\documentclass{SciPost}

\hypersetup{
    colorlinks,
    linkcolor={red!50!black},
    citecolor={blue!50!black},
    urlcolor={blue!80!black}
}

\usepackage[bitstream-charter]{mathdesign}
\usepackage{bm}
\usepackage{placeins}
\usepackage{physics}
\usepackage{cancel}
\usepackage[normalem]{ulem}

\DeclareSymbolFont{usualmathcal}{OMS}{cmsy}{m}{n}
\DeclareSymbolFontAlphabet{\mathcal}{usualmathcal}

\fancypagestyle{SPstyle}{
\fancyhf{}
\lhead{\colorbox{scipostblue}{\bf \color{white} ~SciPost Physics }}
\rhead{{\bf \color{scipostdeepblue} ~Submission }}

\fancyfoot[C]{\textbf{\thepage}}
}

\renewcommand{\i}{\mathop{\mathrm{i}}}

\begin{document}

\pagestyle{SPstyle}

\begin{center}{\Large \textbf{\color{scipostdeepblue}{
Spectral Riemann Sheet Topology of Gapped Non-Hermitian Systems\\
}}}\end{center}

\begin{center}\textbf{
Anton Montag\textsuperscript{1,2$\star$},
Alexander Felski\textsuperscript{1$\dagger$} and
Flore K. Kunst\textsuperscript{1,2$\ddagger$}
}\end{center}

\begin{center}
{\bf 1} Max Planck Institute for the Science of Light, Erlangen, Germany
\\
{\bf 2} Department of Physics, Friedrich-Alexander-Universit\"at Erlangen-N\"urnberg, Erlangen, German
\\[\baselineskip]
$\star$ \href{mailto:email1}{\small anton.montag@mpl.mpg.de}\,,\quad
$\dagger$ \href{mailto:email2}{\small alexander.felski@mpl.mpg.de}\,,\quad
$\ddagger$ \href{mailto:email2}{\small flore.kunst@mpl.mpg.de}
\end{center}

\section*{\color{scipostdeepblue}{Abstract}}
\textbf{\boldmath{%
We show topological configurations of the complex-valued spectra in gapped non-Hermit\-ian systems.
These arise when the distinctive EPs in the energy Riemann sheets of such models are annihilated after threading them across the boundary of the Brillouin zone.
This results in a non-trivially closed branch cut that is protected by an energy gap in the spectrum.
Their presence or absence establishes topologically distinct configurations for fully non-degenerate systems and tuning between them requires a closing of the gap, forming exceptional point degeneracies.
We provide an outlook toward experimental realizations in metasurfaces and single-photon interferometry.
}}

\vspace{\baselineskip}

\noindent\textcolor{white!90!black}{%
\fbox{\parbox{0.975\linewidth}{%
\textcolor{white!40!black}{\begin{tabular}{lr}%
  \begin{minipage}{0.6\textwidth}%
    {\small Copyright attribution to authors. \newline
    This work is a submission to SciPost Physics. \newline
    License information to appear upon publication. \newline
    Publication information to appear upon publication.}
  \end{minipage} & \begin{minipage}{0.4\textwidth}
    {\small Received Date \newline Accepted Date \newline Published Date}%
  \end{minipage}
\end{tabular}}
}}
}


\vspace{10pt}
\noindent\rule{\textwidth}{1pt}
\tableofcontents
\noindent\rule{\textwidth}{1pt}
\vspace{10pt}


\section{Introduction}
\label{sec:intro}
Non-Hermitian systems feature properties with no counterpart in Hermitian models, such as skin states, dissipative phase transitions, and unidirectional transmission~\cite{KatoBook,Heiss2012,Ashida2020}.
These features are closely tied to the topology of the complex-valued spectral structure of such systems~\cite{Bergholtz2021,Budich2019}.
The extensive studies of non-Hermitian spectra have focused by-and-large on so-called exceptional points (EPs)~\cite{Heiss2012,Miri2019,Sayyad2022,Zhen2015,Zhang2019,Zhang2020,Crippa2021,Hodaei2017,Jing2017,Ding2016,Montag2024a,Montag2024b} and spectral winding numbers~\cite{Gong2018,Delplace2021,Leykam2017,Shen2018}.
Here, we instead demonstrate topologically distinct Riemann-sheet configurations of the complex energy function in fully non-degenerate time-reversal-symmetric non-Hermitian systems.
For a two-dimensional periodic two-band system we show that four topologically distinct configurations are realized by closed non-contractible branch cuts in the Riemann sheet structure over the Brillouin zone.
These configurations are structurally analogous to the ground state of the toric code, in which they represent a protected logical qubit pair within a toroidal spin lattice~\cite{Kitaev2003}.
The $\mathbb{Z}_2 \times \mathbb{Z}_2$ topological order that protects the toric code ground states translates to a $\mathbb{Z}_2 \times \mathbb{Z}_2$ topological invariant for the Riemann sheet structure of the non-Hermitian lattice.
We extend this analogy by showing that EPs in the non-Hermitian spectrum emerge similarly to excitations of the toric code~\cite{Kitaev2003,Kitaev2006,Han2007}.
While there are three types of excitations in the conventional toric code, the number of different excitations in the non-Hermitian model is tied to its band structure.
We conclude with an outlook on potential experimental implementations in optical, plasmonic, and mechanical metasurfaces, and single-photon interferometers.

\section{Non-Hermitian Bloch Hamiltonians and spectral Riemann sheets}

Two-dimensional lattice structures can be described by a Bloch Hamiltonian, which is defined over the toroidal surface formed by the two-dimensional Brillouin zone.
Allowing for dissipative processes on the lattice results in a non-Hermitian Bloch Hamiltonian, $H(\bm{k})\neq H^\dagger(\bm{k})$, and complex-valued eigenenergies  $\epsilon(\bm{k})$.
Topology in conventional Hermitian systems operates on the level of eigenstates, whereas in non-Hermitian systems topological structures may also emerge in the Riemann sheets describing the energy spectrum~\cite{Bergholtz2021,Kawabata2019,Delplace2021}.
\begin{figure}[t]
    \centering
    \includegraphics[width=0.79\linewidth]{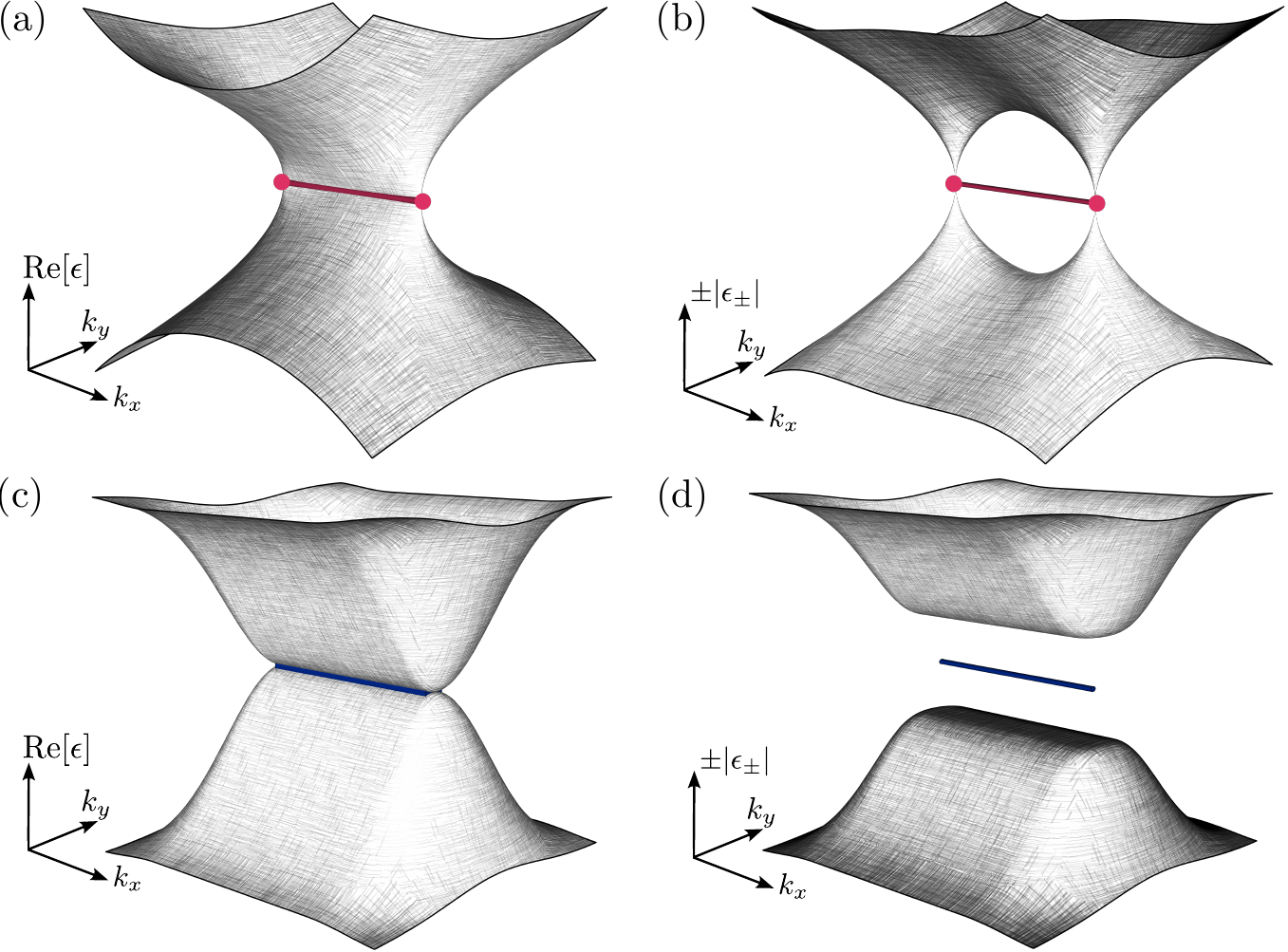}
    \caption{Illustration showing the real part and signed absolute value of the Riemann sheet structure of non-Hermitian two-band spectra: (a), (b) show an EP2 pair connected by a branch cut identified with the Fermi arc in red. This line is referred to as a Fermi cut; (c), (d) show an open Fermi arc due to band touching of $\Re[\epsilon(\bm{k})]$ in blue. The corresponding imaginary part, and $\pm|\epsilon_\pm(\bm{k})|$ accordingly, is gapped along the Fermi arc, shown in (d). In two dimensions the Fermi cut is protected by the presence of the stable EP2 pair, while the Fermi arc may disappear under perturbations.}
    \label{fig:1}
\end{figure}
The most prominent feature in non-Hermitian topology are exceptional points of order $n$ (EP$n$s), which cannot arise in Hermitian systems \cite{KatoBook,Heiss2012,Miri2019,Ashida2020,Bergholtz2021}.
At these points the Hamiltonian becomes non-diagonalizable, due to the coalescence of $n$ eigenvectors, and the spectrum exhibits an $n$-fold degeneracy. In the energy sheet structure, EPs manifest as branch points , cf. Fig.~\hyperref[fig:1]{1(b)}.
Their classification based on topological charges tied to the local enveloping spectral structure has been the focus in studies of generic non-Hermitian Hamiltonians and non-Hermitian systems with (anti-) unitary symmetries in particular~\cite{Leykam2017,Shen2018,Delplace2021,Sayyad2022,Montag2024a}.  

The prevalent exceptional structure in two-dimensional parameter spaces are EP2s.
Consider a two-band system, described by a $2\times2$ Bloch Hamiltonian, which gives rise to two complex-valued energy sheets over the Brillouin zone.
In this two-dimensional space, EP2s generically appear in pairs, see Fig.~\hyperref[fig:1]{1(a-b)}, with opposite spectral winding around the respective EP2 defining a topological charge~\cite{Leykam2017,Shen2018}.
Any such pair is connected by a Fermi arc, an imaginary Fermi arc, and a branch cut of the Riemann sheets \cite{Berry2004,Kozii2017,Zhou2018,Bergholtz2021,Wang2024b,Cheng2026}: Fermi arcs are lines on which the real  part of the eigenenergies is degenerate, while along imaginary Fermi arcs the imaginary part of the eigenenergies is degenerate.
Note that such arcs can also exist independent of EPs, for example at band touching lines, see Fig.~\hyperref[fig:1]{1(c)}.
A branch cut, on the other hand, is a line along which a continuous multi-sheeted covering of the Brillouin zone is separated into well-defined single-valued sheets; the path of such branch cuts can be chosen freely in general.
Here we choose the common identification of the branch cuts with the Fermi arc that always connects the EP2 pair, and refer to the combined object as a \emph{Fermi cut}~\cite{footnote1}.
The presence of this Fermi cut is protected and results in a continuously connected energy structure.
The protection originates from the presence of the stable EP2 pair and prevents the Fermi cut from being removed by any small perturbation.
While such stable features of non-Hermitian systems have been studied, the presence of EPs implies spectral degeneracies that prevent the definition of a conventional topological gap.

Consider the following toy model as an example:
\begin{equation}
    H(\bm{k}) = \begin{pmatrix}
        \sqrt{3}\cos  k_x & \sqrt{3}\sin k_x + \i +\sqrt{2}\,\alpha \cos k_y \\
        \sqrt{3}\sin k_x + \i - \sqrt{2}\, \alpha \cos k_y & - \sqrt{3}\cos k_x
    \end{pmatrix} \, , \qquad \alpha\in\mathbb{R}^+_0 \, ,
\end{equation}
with the spectral Riemann sheets described by 
\begin{equation}
    \epsilon_\pm(\bm{k}) = \pm\sqrt{2[1-\alpha^2\cos^2 k_y + \sqrt{3}\i \sin k_x]} \, .
\end{equation}
For $\alpha<1$ EPs are absent, and the spectrum is gapped, meaning $|\Delta\epsilon(\bm{k})|>0$.
At $\alpha=1$ the two complex-valued energy sheets touch at the points 
$(k_x,k_y)\in\{(0,0),(0,\pi),(\pi,0),(\pi,\pi)\}$.
These band-touching points then split into pairs of EPs for $\alpha>1$.
For instance, the EP2 pair originating at $(k_x,k_y)=(0,0)$ will be located at 
$\text{EP}_\pm =(0,\pm\arccos \alpha^{-1} )$ 
and connected by a Fermi cut along the curve 
$\mathcal{C}_\text{FC}:\{(0,k) \,\vert\, k\in(-\arccos \alpha^{-1},\arccos \alpha^{-1}) \}$.
In Figs.~\hyperref[fig:1]{1(a-b)} the real part and the absolute value of the complex-valued spectrum for just this model were shown with $\alpha=\sqrt{2}$.
By perturbing the Hamiltonian the path of this Fermi cut may be deformed, but it cannot be removed since its presence is protected by the EPs under any  perturbation that does not lead to a local recombination of the EPs.

We reiterate that in the presence of EPs the complex energy gap is closed.
However, our aim is a spectral classification of \emph{gapped} non-Hermitian systems. 
To this end, we investigate the behavior of the Fermi cuts when the EPs are merged again.

\section{Gapped Riemann sheet topology}

In the following we develop a topological classification of two-dimensional gapped non-Hermitian spectra based on \emph{non-contractible} stable Fermi cuts.
We begin by discussing the persistence of Fermi cuts after the merger of EPs in periodic parameter spaces.
Their topological protection is founded in the complex-valued band gap. 
This stability of Fermi cuts is contrasted with the fragility of non-contractable Fermi arcs under infinitesimal perturbation of the Hamiltonian.
In addition, we demonstrate the reduction of this topological classification for two-dimensional gapped non-Hermitian two-band systems to a $\mathbb{Z}_2 \times \mathbb{Z}_2$ topological invariant in the presence of time-reversal symmetry.

\subsection{Closed Fermi cuts}

To meaningfully address closed Fermi cuts in gapped systems, a definition in the absence of EPs is required.
For the creation of a stable EP pair the presence of a closed Fermi arc or imaginary Fermi arc is a prerequisite.
An EP pair emerges on this arc, opening it and creating the complementary arc between the pair; for example, if an EP pair emerges on a Fermi arc an imaginary Fermi arc is created between the pair of EPs.
In addition, a branch cut connects the EP pair.
We can now use these properties for a definition of Fermi cuts in non-periodic parameter spaces: A Fermi cut is an open branch cut that is identified with a Fermi arc, which is closed by an imaginary Fermi arc.
If either the real or the imaginary Fermi arcs are closed, their counterpart vanishes and the Fermi cut depreciates to a Fermi arc.

In periodic parameter spaces, such as the toroidal Brillouin zone, we have to distinguish between EPs emerging on contractible or non-contractible (imaginary) Fermi arcs to define Fermi cuts. 
As long as the EP pair is created on a contractible (imaginary) Fermi arc, the previous definition suffices.
If EPs emerge on non-contractible Fermi arcs, however, the discussion is more nuanced.
It is important to note that non-contractible (imaginary) Fermi arcs can only appear pairwise without closing the complex energy gap. 
Yet, such a pair of non-contractible (imaginary) Fermi arcs can always be deformed continuously to a single contractible (imaginary) Fermi arc.
For the creation of EP pairs, let us consider the example of two non-contractible closed imaginary Fermi arcs in the following.
As before, creating an EP pair on one of these arcs results in an open Fermi arc, which is closed by an open imaginary Fermi arc in a non-contractible manner, while the other non-contractible imaginary Fermi arc remains.
Thus, the union of these three arcs combines to a contractible loop.
This remains true even when the EPs are merged such that the open imaginary Fermi arc vanishes; the Fermi arc is now closed in a non-contractible manner, however, its union with the remaining non-contractible imaginary Fermi arc still combines to a contractible loop.
From these considerations we obtain a definition of a Fermi cut in a periodic parameter space without invoking EPs directly: a Fermi cut is either an open or non-contractible closed branch cut that is identified with a Fermi arc, whose union with an imaginary Fermi arc forms a contractible closed loop.
This definition includes the Fermi cuts spanned between a pair of EPs, as well as the Fermi cut defined by a non-contractible closed Fermi arc running parallel to the associated imaginary Fermi arc around the periodic parameter space.

To construct the topological classification of non-Hermitian spectra, we utilize the topological protection of Fermi cuts.
For the creation of a closed Fermi cut, as described above, a pair of EP2s must be generated and pulled apart, requiring a closing of the complex energy gap.
To reopen the complex energy gap the EP2 pair has to be recombined, thus also closing the Fermi cut.
If one of the EPs is threaded through the full Brillouin zone and \emph{across} the periodic boundary, the Fermi cut survives even after the EPs are merged again.
This results in a fully non-degenerate system with a finite gap, which prevents the perturbative creation of EP pairs, thus protecting the Fermi cut.
The resulting closed branch cut runs through the whole Brillouin zone along a non-contractible loop.
The threading of the EP changes the structure of the spectral Riemann sheets non-trivially, resulting in a fully non-degenerate continuously-connected energy structure.
We stress that it is the toroidal topology of the Brillouin zone that facilitates such a closed Fermi cut.

\begin{figure}
    \centering
    \includegraphics[width=\linewidth]{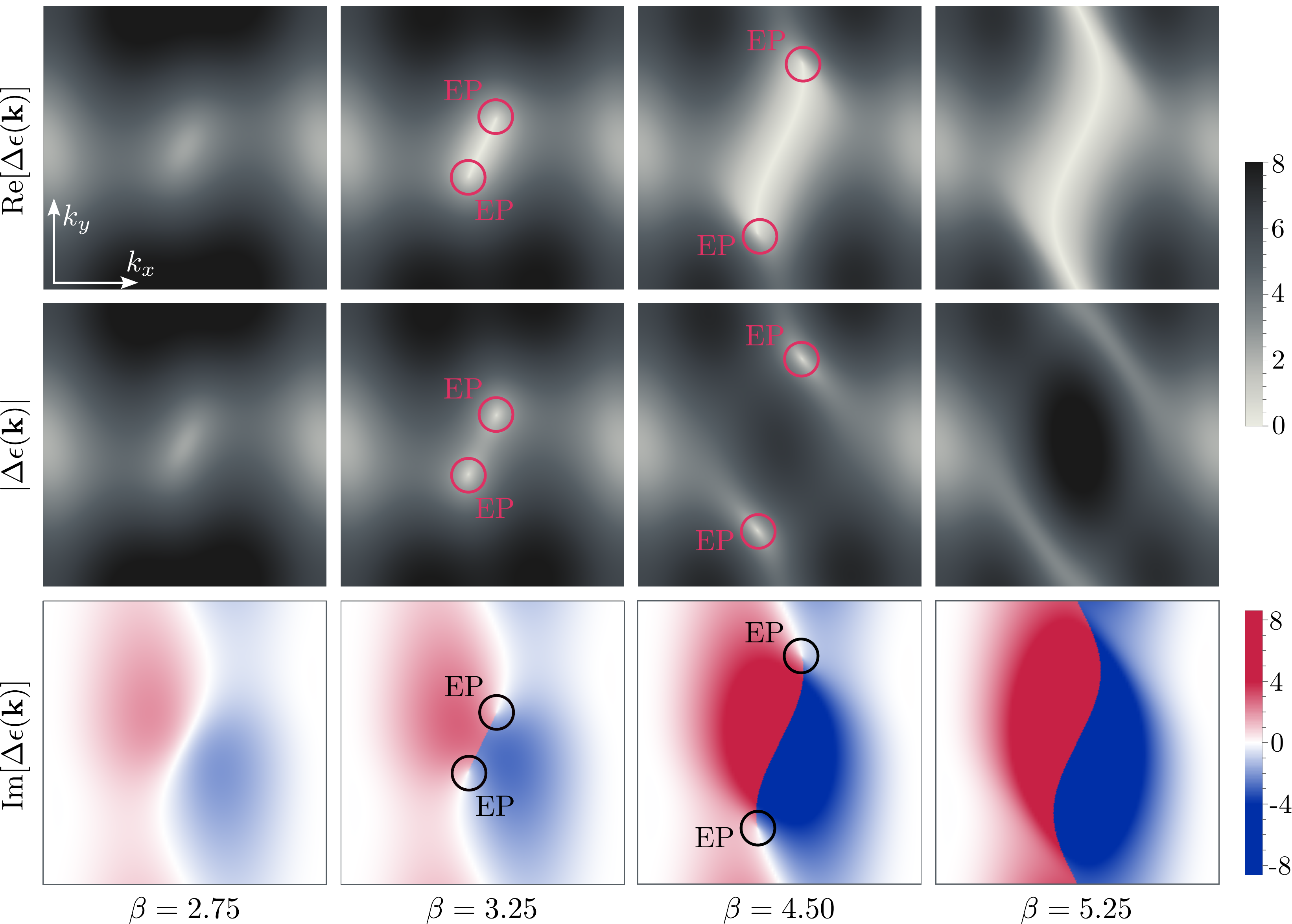}
    \caption{Creation of a closed Fermi cut. Shown are the real part, the absolute value and the imaginary part of the difference between the eigenenergies $\Delta \epsilon(\bm{k})=\epsilon_+(\bm{k})-\epsilon_-(\bm{k})$ for four different values of the tuning parameter $\beta$ over the full Brillouin zone. The eigenenergies $\epsilon_\pm(\bm{k})$ are labeled according to the sign of the real part. The two intermediate values of $\beta$ correspond to gapless spectra, where EPs are highlighted by the red/black circles. In the first row, the creation of the closed Fermi cut is clearly visible, with the last plot showing the persistence of the Fermi cuts after the merger of the EPs. The jump in the imaginary part of the energies across the Fermi cut is apparent.}
    \label{fig:threading}
\end{figure}

To illustrate this procedure consider the non-Hermitian Bloch Hamiltonian
\begin{equation}
    H(\bm{k}) = \scalebox{0.9}{$\begin{pmatrix}
        3+\cos k_x-\cos k_y &  (-1+\i)\sin k_x+\tfrac{1}{2}\sin k_y+i \tfrac{\beta}{2}(1+\cos k_x)\\
        (-1-\i)\sin k_x+\frac{1}{2}\sin k_y+i \tfrac{\beta}{2}(1+\cos k_x) & -3-\cos k_x+\cos k_y
    \end{pmatrix}$} ,
\end{equation}
with $\beta\in\mathbb{R}^+_0$.
The corresponding spectrum is gapped for $\beta<3$ and $\beta>5$: $|\Delta \epsilon(\bm{k})|>0$, where $\Delta \epsilon(\bm{k})$ denotes the difference between the two eigenenergies. 
At $\beta=3$ a pair of EP2s is created at $(k_x,k_y)=(0,0)$ and separated with increasing $\beta$, resulting in a connecting open Fermi cut.
At $\beta=5$ the EPs merge across the Brillouin zone boundary, reinstating a gapped Riemann energy sheets. 
In contrast to a local recombination, however, in this global approach the Fermi cut remains, which results in a single-sheeted Riemann sheets on which all eigenenergies can be connected via continuous curves.
The process of threading the EPs is shown in Fig.~\ref{fig:threading}.  

\begin{figure*}
    \centering
    \includegraphics[width=\linewidth]{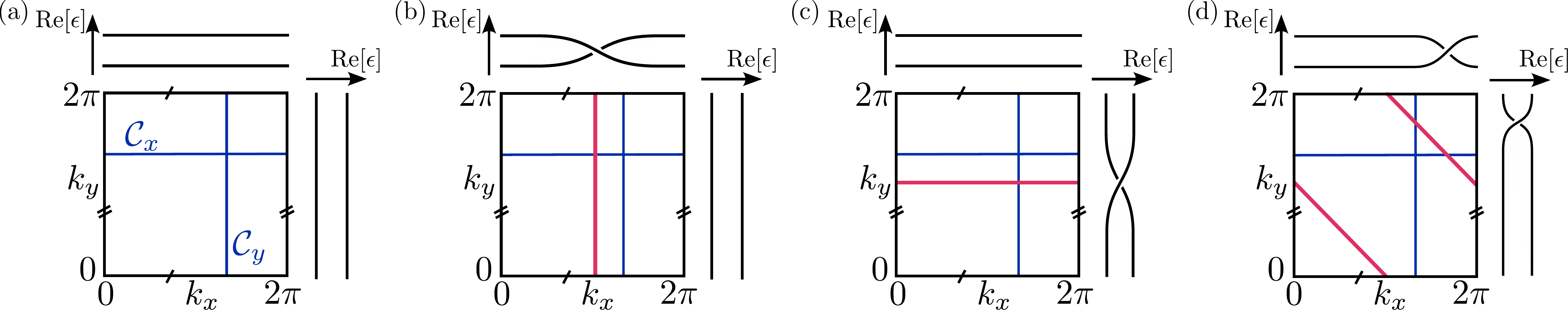}
    \caption{We show the three distinct non-contractible Fermi cuts possible in the two-dimensional Brillouin zone, where Fermi cuts are represented by red lines. Two distinct non-contractible loops, $\mathcal{C}_x$ and $\mathcal{C}_y$, around the different holes of the torus are indicated in blue. Along these loops the eigenenergy braids, shown next to the Brillouin zone, are projections of the continuous complex-valued gapped spectra. They can be measured to determine the topological invariants.}
    \label{fig:non_contract_loops}
\end{figure*}

The topological protection of the Fermi cuts after threading and merging the EPs can be understood from the perspective of eigenenergy braids.
These braids are defined on closed lines around the holes of the toroidal Brillouin zone, cf. Fig.~\ref{fig:non_contract_loops}.
A Fermi cut results in a crossing of the braids orthogonal to the cut.
Perturbations cannot change the non-trivial braids of single Fermi cuts, ensuring their stability~\cite{Wojcik2022}.
This is similar to the braid protection of third-order EPs observed in Ref.~\cite{Koenig2023}.
In contrast to previous works, which consider braids along loops that encircle the full Brillouin zone,  the topological protection here arises from braids along non-contractible loops.
Unlike Ref.~\cite{Koenig2023}, where only isolated higher-order EPs result in non-trivial braids, this allows to capture non-contractible Fermi cuts.

The presence or absence of non-trivially closed Fermi cuts results in a different connection of the Riemann energy sheets.
This connectedness describes the (im-)possibility to change between sheets by following a non-contractible loop $\mathcal{C}$ in the Brillouin zone. 
If the eigenenergy crosses an even number of Fermi cuts, it returns to its initial value after traversing the parametric loop $\mathcal{C}$--the sheets are disconnected along this direction.
When crossing an odd number of Fermi cuts, the sheets are connected instead and the eigenenergy does not return to the initial value.
By distinguishing whether the Riemann sheet is connected along both directions, we define four distinct configurations of the system.
These configurations are only well-defined for a gapped non-degenerate system without EPs present.
Tuning the system from one of these states to another thus requires the merger of all EPs after the threading procedure.
To distinguish the configurations we define two topological invariants, which are derived from the permutations of the eigenenergies along non-contractible loops through the Brillouin zone.
The invariant $m_\alpha=0$ is associated with non-connected and $m_\alpha=1$ with connected Riemann sheet structures, where $\alpha\in\{x,y\}$ indicates whether the path $\mathcal{C}$ runs along the $k_x$ or $k_y$ direction.
Specifically, a non-vanishing invariant $m_\alpha$ corresponds to a non-trivial permutation $\pi_\alpha$ of the two energies, that is, exchanging the eigenvalues $\epsilon_1$ and $\epsilon_2$ of the two eigenenergy sheets:
\begin{equation}
    m_\alpha = 1-\delta_{\pi_\alpha(1),1} \, , \label{eq:inv}
\end{equation}
in terms of the Kronecker delta $\delta_{i,j}$.
This exchange of the eigenenergies is reminiscent of the discussion of state switching in Ref.~\cite{Ryu2025}.
For example, if a non-contractible loop in $k_\alpha=k_x$ direction crosses an odd number of Fermi cuts, the eigenenergies exchange and we find $\pi_x(1)=2$, and thus $m_x=1$.

\subsection{Snapping Fermi arcs}
To highlight the topological protection unique to closed Fermi cuts, we instead consider a single Fermi arc running along a non-contractible loop.
We stress that the presence of Fermi arcs does not change the topology of the spectral Riemann sheets.
Rather these Fermi arcs are the result of a finely tuned band-touching of the real part of the Riemann energy sheets.
Contrary to the Fermi cuts discussed above, the non-contractible Fermi arcs are fragile and snap under infinitesimal perturbation, that is the full arc can be instantaneously removed.
To illustrate this absence of protection, let us consider the Bloch Hamiltonian 
\begin{equation}
    H_\text{FA} (\bm{k}) = \begin{pmatrix}
        \delta+i & \left(1-\cos k_x\right) + \frac{i}{2} \\
        \left(1-\cos k_x \right) + \frac{i}{2} & -\delta-i
    \end{pmatrix},
\end{equation}
which realizes a non-contractible Fermi arc along the $k_y$-direction of the Brillouin zone for $\delta=0$.
Any non-vanishing $\delta \in \mathbb{R}^+$ does not continuously deform the Fermi arc, but abruptly snaps it.
Despite the presence of a gapped spectrum, the Fermi arc is not protected, because it can be removed without closing the gap.
This is a consequence of the Fermi arc being a spectral observable only.
The eigenenergies remain separated into distinct sheets even in the presence of a Fermi arc, compare also Fig.~\hyperref[fig:1]{1(c-d)}.
Such a band touching is not topologically protected, and therefore already destabilized by small perturbations.

\subsection{Time-reversal-symmetric non-Hermitian systems}

In principle, the generation of closed Fermi cuts can be iterated:  
By generating multiple pairs of EP2s and threading them through the Brillouin zone, many topologically distinct gapped Riemann energy sheets configurations, with higher braid indices analogous to Fig.~\ref{fig:non_contract_loops}, can be realized.
Yet, these models are usually fine-tuned, with their realization requiring various long-range interactions.
However, in the special case of time-reversal-symmetric non-Hermitian models, satisfying~\cite{Kawabata2019} 
\begin{equation}\label{eq:TR}
    \mathcal{T}H^*(\bm{k})\mathcal{T}^{-1} = H(-\bm{k}) , \quad \text{with} \;\; \mathcal{T}\mathcal{T}^\dagger=\mathbb{1},
\end{equation}
a simple and robust topological classification of gapped Riemann energy sheets can be found.

Time-reversal symmetry does not induce EPs, because it does not constrain the spectrum locally, but instead relates the spectrum at each point $\bm{k}$ in parameter space to the spectrum at the time-reversed momentum $-\bm{k}$~\cite{Kawabata2019,Sayyad2022}.
However, this restricts the points at which single EP2 pairs can emerge generically, because they have to satisfy the symmetry~\cite{Stalhammar2025}.
They can now only be generated and merged at time-reversal invariant momenta, that is at the points $(k_x,k_y)\in\{(0,0),(0,\pi), \allowbreak(\pi,0), \allowbreak (\pi,\pi)\}$ in the two-dimensional Brillouin zone.
As an important consequence, non-contractible Fermi cuts can only cross the Brillouin zone boundary at these points.
The underlying non-trivial energy braids, shown in Fig.~\ref{fig:non_contract_loops}, moreover imply the existence of a non-contractible imaginary Fermi arc, parallel to the real Fermi arc defining the path of the Fermi cut.
Given the time-reversal symmetry of the model, this imaginary Fermi arc is required to cross the Brillouin zone boundary at a time-reversal-invariant momentum as well.
In addition, it cannot cross the boundary at the same point as the Fermi cut, which would imply the presence of an EP2: if a Fermi cut in the $x$-direction crosses at $(k_x,k_y)=(\pi,0)$ the imaginary Fermi arc must cross at $(k_x,k_y)=(\pi,\pi)$ and vice versa.

\begin{figure}[!]
    \centering
    \includegraphics[width=\linewidth]{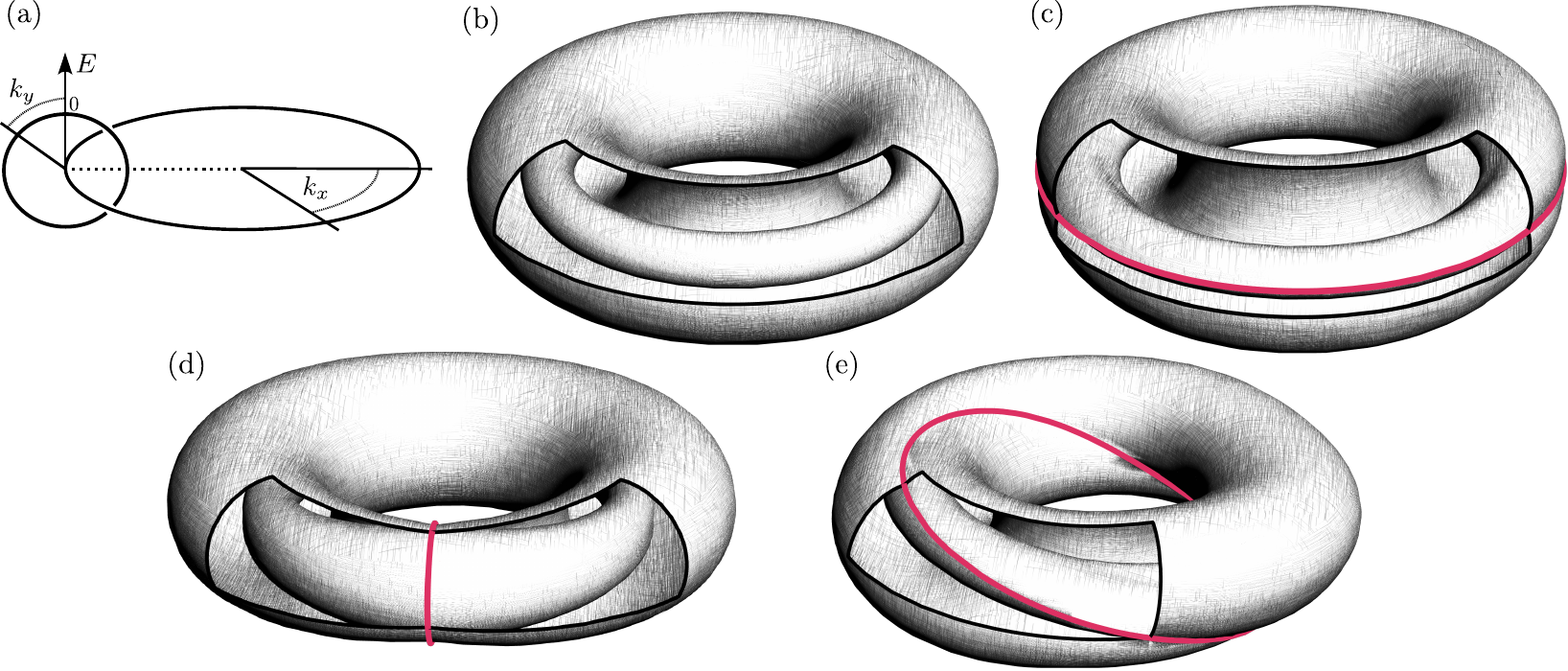}
    \caption{The real part of the spectral Riemann sheets over the Brillouin zone is shown for the topologically distinct configurations. (a) Coordinate system of the toroidal Brillouin zone, where the two quasi-momenta $k_x,k_y$ correspond to the two angles in toroidal coordinates and the radius encodes the real part $E$ of the complex-valued spectrum; (b)-(e): Topologically distinct fully non-degenerate configurations of the non-Hermitian model $H_\text{FC}(\bm{k})$ analogous to the ground states of the toric code. Sections of the sheet with larger real part of the eigenvalues are removed to show the lower sheet and highlight how the non-contractible Fermi cuts (red lines) appear as the intersections in the real part of the energy sheets. The configurations are distinguished by the direction along which the non-contractible Fermi cuts are closed around the toroidal Brillouin zone, which is quantified by the topological invariants $(m_x,m_y)$: (b) $(0,0)$, (c) $(0,1)$, (d) $(1,0)$, (e) $(1,1)$. If $m_\alpha=0$ no Fermi cut is crossed by traversing the spectrum along a non-contractible closed loop in this direction and $m_\alpha=1$ corresponds to crossing one Fermi cut.}
    \label{fig:models}
\end{figure}

Due to these additional constraints on the position of the Fermi cut and the associated imaginary Fermi arc, generic time-reversal-symmetric non-Hermitian systems admit only a single Fermi cut along each direction.
Any additional Fermi cut results either in an EP2 emerging at a time-reversal invariant momentum on the boundary, thus closing the complex energy gap, or it combines with the first Fermi cut, forming a contractible loop.
Therefore, gapped time-reversal-symmetric non-Hermitian systems are characterized by only four distinct configurations of the Riemann energy sheets.
They are distinguished by the presence or absence of a single non-contractible Fermi cut along either direction in the Brillouin zone, and uniquely labeled by their respective connectedness in Eq.~(\ref{eq:inv}). 
For the time-reversal symmetric models the topological invariants can be formulated in terms of the discriminant of the Bloch Hamiltonian, 
\begin{equation}
    \Delta(\bm{k}) = \{\Tr[H(\bm{k})]\}^2 - 4 \det[H(\bm{k})] \, ,
\end{equation}
which has the property $\Delta(\bm{k})=\Delta^*(-\bm{k})$.
At the time-reversal invariant momenta the discriminant is real and non-vanishing for gapped spectra, so that its sign is well defined.
Therefore we can define the topological invariants characterizing the four distinct configurations as
\begin{align}
    (-1)^{m_x} &= \textrm{sign}[\Delta(\pi,0)/\Delta(0,0)] = \textrm{sign}[\Delta(0,\pi)/\Delta(\pi,\pi)] \, , \\
    (-1)^{m_y} &= \textrm{sign}[\Delta(0,\pi)/\Delta(0,0)] = \textrm{sign}[\Delta(\pi,0)/\Delta(\pi,\pi)] \, .
\end{align}
The  advantage of this purely algebraic formulation lies in its immediate connection to the Bloch Hamiltonian.
For representative cases of the topologically distinct configurations, the real part of the Riemann energy sheets over the toroidal Brillouin zone is shown in Fig.~\ref{fig:models}.
The non-contractible Fermi cuts, across which the different sheets are connected, are highlighted in red.

\section{Time-reversal-symmetric non-Hermitian spectra and the toric code ground states}

The $\mathbb{Z}_2 \times \mathbb{Z}_2$ topological invariant, see Eq.~(\ref{eq:inv}), characterizing gapped time-reversal-symmetric non-Hermitian two-band systems is reminiscent of the topological order governing the ground state manifold of the toric code.
We briefly recall the toric code Hamiltonian and its ground states, before introducing a time-reversal-symmetric non-Hermitian Hamiltonian, which realizes four analogous topologically distinct Riemann energy sheet configurations.
A further analogy between EPs and excitations in the toric code is discussed, before proceeding to generalize the findings to non-Hermitian multi-band systems.

\subsection{Toric code ground states}

\begin{figure*}
    \centering
    \includegraphics[width=0.9\linewidth]{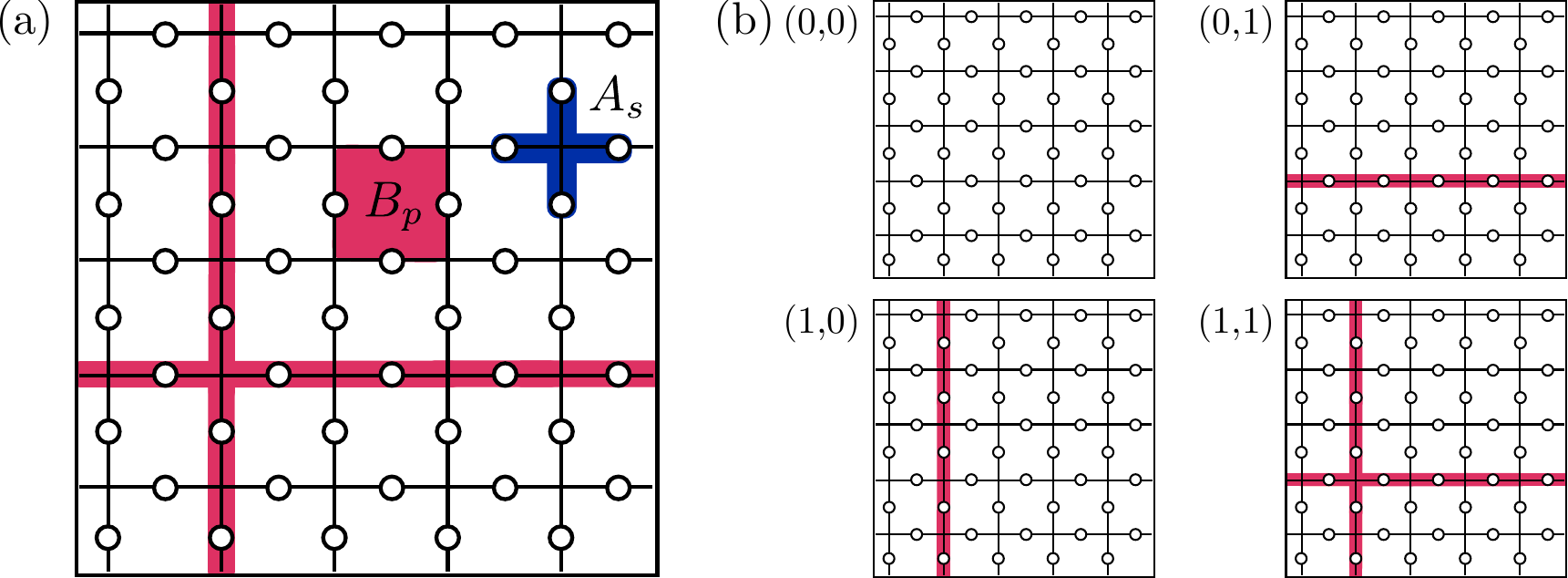}
    \caption{Illustration of the toric code on a square lattice given periodic boundary conditions. Physical qubit degrees of freedom are represented by white circles. (a) A star and a plaquette operator, $A_s$ and $B_p$, are shown in blue and red, respectively. Non-contractible loops of flipped spins, which define the ground states, are indicated by red lines. (b) Representatives of the four topologically distinct ground states are shown. They are distinguished by the different possible non-contractible closed loops on the toroidal lattice.}
    \label{fig:toric_code}
\end{figure*}

A prominent application of topology is the protection of the ground states in Kitaev's toric code~\cite{Kitaev2003}.
These states implement two logical qubits within a square lattice that is defined on a toroidal surface and has physical qubit degrees of freedom on its links.
The key idea of this encoding is the definition of a Hamiltonian in terms of overlapping but commuting local operators.
Due to the overlap, such a model and its states are highly non-trivial.
The states can be fully determined nevertheless because all local operators commute.
For the conventional toric code, this can be achieved by defining the star and plaquette operators,
\begin{align}
    A_s = \prod_{i\in s} \sigma_i^x \quad \text{and} \quad B_p = \prod_{i\in p} \sigma_i^z  \label{eq:stabelizers} \, ,
\end{align}
where the $\sigma_i^\alpha$ with $\alpha\in\{x,y,z\}$ are Pauli operators acting on the physical qubits.
Here products run over the links coming together at a lattice site $s$, or over the boundary links of a plaquette $p$; cf. Figs.~\hyperref[fig:toric_code]{5(a)}.
In these terms, the toric code is governed by the Hamiltonian
\begin{equation}\label{eq:toric_code_H}
    H_\text{tc} = - \sum_s A_s - \sum_p B_p \, ,
\end{equation}
summing over all sites and plaquettes.

Measuring a plaquette or star operator yields $\pm1$ and does not affect the model, since $[H_\text{tc},A_s]=[H_\text{tc},B_p]=0$.
Thus $A_s$ and $B_p$ form a basis for error detection making them so-called stabilizers~\cite{Kitaev2003,Shor1995}.
A state for which all of these stabilizers are measured to be $+1$ minimizes the energy of $H_\text{tc}$ making it a ground state of the system.
The ground state is not unique, which can be seen by considering the following.
If the physical qubits are measured in the $\sigma^x$ basis, the configuration with only spin-up states achieves $A_s=+1$.
So does any configuration that can be obtained from flipping spins along closed loops.
Projecting these configurations onto the subspace with $B_p=+1$ results in the ground-state mani\-fold of the toric code.
Any contractible loop of flipped spins can be removed by applying plaquette operators, which leave the ground-state manifold unaffected while flipping all the physical qubits around a plaquette.
Therefore the highly degenerate set of ground-state configurations is classified by four topologically distinct realizations of flipped spins along \emph{non-contractible} loops on the toroidal surface, see Fig.~\hyperref[fig:toric_code]{5(b)}.
As such, they encode a logical two-qubit system that is protected by the topology of the non-contractible loops.
The presence or absence of each possible non-contractible loop of flipped spins defines a single logical qubit.

\subsection{Analogous time-reversal-symmetric non-Hermitian system}

Based on the four configurations of time-reversal-symmetric non-Hermitian spectra distinguished by the combinations of topologically protected closed Fermi cuts, we construct the analogy to the ground-state manifold of the toric code in an exemplary model.
Any closed Fermi cut is mapped to a non-contractible loop of flipped spins in the toric code, and the four different ground states are represented by $(m_x,m_y)\in\{(0,0),(1,0),(0,1),(1,1)\}$.
The ground-state manifold is realized by the tunable non-Hermitian system with the Bloch Hamiltonian
\begin{equation}\label{eq:nH_toric}
    H_\text{FC}^{(m_x,m_y)}(\bm{k}) =
    \begin{pmatrix}
        3-\cos k_x\!-\cos k_y & \sin s_{\bm{k}}  \!- 3i \!\left(1\!-\!\cos s_{\bm{k}}\right)\\[2pt]
        \sin s_{\bm{k}} \! - 3i \!\left(1\!-\!\cos s_{\bm{k}}\right) & -3+\cos k_x\!+\cos k_y
    \end{pmatrix} \, ,
\end{equation}
where $s_{\bm{k}}=m_x \,k_x+ m_y \, k_y$.
The model obeys time-reversal symmetry with respect to the unitary transformation
\begin{equation}
    \mathcal{T} H_\text{FC}^{(m_x,m_y)}(\bm{k}) \mathcal{T}^{-1} = \left[H_\text{FC}^{(m_x,m_y)}(-\bm{k})\right]^*, \qquad \mathcal{T} = \sigma_z.
\end{equation}
Figure~\ref{fig:models} showcases the four topologically distinct configurations of the non-Hermitian toric-code analog and highlights the non-trivially closed Fermi cuts, which run along the curves defined by $s_{\bm{k}} = (\pi \; \text{mod} \; 2\pi)$ for the given system.
When tuning continuously between two topologically distinct configurations, $H(\lambda) = (1-\lambda) H_\text{FC}^{(m_x,m_y)} + \lambda H_\text{FC}^{(m'_x,m'_y)}$ with $\lambda\in[0,1]$ and $(m_x,m_y)\neq(m'_x,m'_y)$, the complex gap closes at a value $\lambda_-$ and opens only after a finite interval at $\lambda_+>\lambda_-$.
Within the interval $(\lambda_-,\lambda_+)$ a pair of EP2s is threaded through the Brillouin zone, changing the topological structure of the Riemann sheets.

In this analogy we capture the topological classification of the toric code defined in terms of non-contractible loops around a torus.
The defining feature of the topologically distinct configurations are non-contractible closed defect lines, which are lines of flipped spins in the toric code and Fermi cuts in the non-Hermitian analogy.
An important distinction is that in the toric code the topological states are ground states of a single Hamiltonian, while we discuss the topologically distinct spectra of different non-Hermitian Bloch Hamiltonians. 

\subsection{Exceptional points as excitations and multi-band generalizations}

The topologically distinct configurations of the non-Hermitian model are only defined for non-degenerate systems in which no EPs are present.
In the following, we show that EPs can be regarded as excitations, analogous to the role of defective sites and plaquettes in the toric code.

The excitations in the conventional toric code are sites $s$ or plaquettes $p$, whose respective stabilizers, $A_s$ or $B_p$, are measured to be $-1$~\cite{Kitaev2003}.
This raises the energy of the system above the lowest possible energy state in which all stabilizers are $+1$.
By flipping a single physical qubit in the $\sigma^x$-basis, a pair of star defects called electric charges $e$ is created on the adjacent sites.
Similarly, flipping in the $\sigma^z$-basis creates two plaquette defects, called magnetic charges $m$, on the plaquettes that share the edge.
These excitation pairs can be pulled apart and moved around the lattice.
They stay connected by a so-called flux line, which is an open line of flipped spins.
Excitations of the same type annihilate when brought together, and the flux lines become closed loops, so that the system returns to a ground state; nontrivial flux threading tunes the system between different states in the ground-state manifold.
Alike charges exchange bosonically, while different charges commute mutually anyonic.
Bringing two different charges together results in a dyon, which is a fermionic composite excitation.

In the non-Hermitian toric-code analog the topologically distinct configurations are characterized by a finite gap.
Transitioning between these configurations requires the spectral gap to close in the form of EPs.
Similar to excitations in the conventional toric code, they appear pairwise and are connected by an open Fermi cut, which plays the role of a flux line.
In this sense, EPs can be interpreted as excitations of the non-Hermitian system.
For a two-band model, the only possible types of excitation are therefore EP2s.
Moreover, the EP2s carry a topological charge $\nu$, given by the spectral winding number 
\begin{equation}
    \nu=-\oint_\mathcal{C} \frac{d\bm{k}}{2\pi}\cdot\nabla_{\bm{k}} \textrm{arg}\left[\Delta \epsilon(\bm{k})\right]=\pm\frac{1}{2} \, ,
\end{equation} 
where $\mathcal{C}$ is a closed path encircling the EP2 and $\Delta \epsilon(\bm{k})$ is the difference between the energy sheets that coalesce at the EP2~\cite{Leykam2017,Shen2018}.
As a result, excitations in the non-Hermitian system carry additional structure compared to the conventional toric code, because only EP2s of opposite topological charge can annihilate.
Combining EP2s of identical charge, on the other hand, results in an unstable EP2 composite with additively combined topological charge.
We stress that, while the pair-wise created EP2s act similarly to the excitations of the toric code in that they transition between topologically distinct configurations, they do not possess the anyonic properties of said excitations. 
The EPs remain classical degeneracies in the spectra of non-Hermitian systems and are not to be confused with the topological excitations of the many-body toric code Hamiltonian.
\begin{figure}
    \centering
    \includegraphics[width=\linewidth]{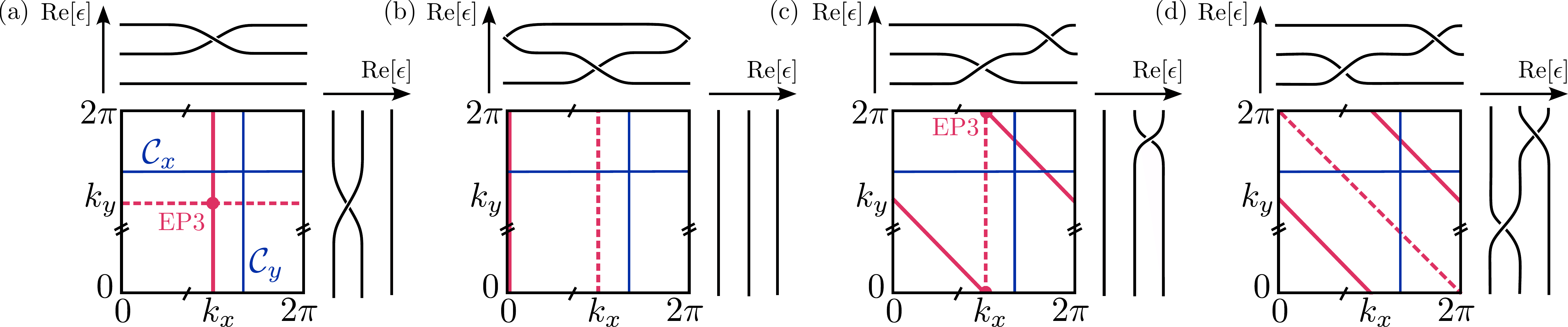}
    \caption{Four representative combinations of non-contractible Fermi cuts (red) possible in non-Hermitian three-band systems. The solid (dashed) lines correspond to Fermi cuts between the pair of Riemann sheets with largest (smallest) real part.
    Intersections of dashed and solid Fermi cuts result in EP3s (red dots), which close the complex energy gap. 
    Two distinct non-contractible loops in the two-dimensional Brillouin zone, $\mathcal{C}_x$ and $\mathcal{C}_y$, along which the eigenenergy braids can be measured to determine the topological invariants, are indicated in blue.}
    \label{fig:multiband}
\end{figure}

In contrast to the conventional toric code, a time-reversal-symmetric non-Hermitian two-band model, while realizing four distinct non-degenerate configurations, allows for only one instead of three excitation types.
The implementation of multiple distinct excitations requires systems with additional bands in the Bloch Hamiltonian.
Adding a third band, for instance, results in three types of EP2s, which are the different possible branch points between pairs of eigenenergy sheets.
When bringing two different EP2s together, EP3s emerge as composite excitations~\cite{Zhong2020,Wiersig2022}.
In general, extending this to $n$-band systems allows for $\sum_{j=2}^n \binom{n}{j}=2^n-(n+1)$ different types of excitations.
These are again associated with topological charges, determined by the winding of the eigenenergies around them \cite{Wojcik2020,Li2021}.

At the same time, added bands affect the possible topologically distinct configurations of the time-reversal-symmetric non-Hermitian multi-band model, which comprise any fully non-degenerate spectral structure.
The number of topologically distinguishable Riemann sheet structures increases, because additional non-contractible Fermi cuts along the $k_x$ and $k_y$ direction are possible between any two bands.
However, intersecting Fermi cuts induce EP3 degeneracies if a single energy sheet is part of both cuts.
Therefore not all combinations of Fermi cuts result in gapped non-Hermitian spectra.
Illustrative examples for different combinations of Fermi cuts, and their respective eigenenergy braids along non-contractible loops $\mathcal{C}_\alpha$ with $\alpha=x,y$, are shown in Fig.~\ref{fig:multiband}.
A set of topological invariants can be defined for each non-degenerate Riemann sheet structure by generalizing Eq.~(\ref{eq:inv}) to all pairs $(p,q)$ of energy sheets and repeated application of the permutation $\pi_\alpha$:
\begin{equation}
    m_\alpha^{pq} = 1 - \delta\Bigl[\, \sum_{j=1}^{n-1} \delta\left[p-\pi_\alpha^j(q)\right]\Bigr] \, ,
\end{equation}
where $\delta[n]$ denotes the Kronecker delta $\delta_{0,n}$.
This establishes the topologically distinct non-degenerate spectral surfaces as a subspace of the space with a $\left(\mathbb{Z}_2 \times \mathbb{Z}_2\right)^m$ topological invariant, where $m=\sum_{j=1}^{n-1}j=\tfrac{1}{2}(n-1)n$.

Overall, extending the non-Hermitian model to multi-band systems goes beyond the initial analogy to the toric code and allows for the realization of multiple distinct excitations. 

\section{Implementation on metasurfaces and in single-photon interferometry}
The implementation of different topologically protected configurations requires a high degree of parametric tunability over a toroidal parameter space.
Measuring the connectedness of the Riemann sheet structure on the other hand is achievable, for example by sampling the real part of the spectrum along non-contractible loops in the Brillouin zone.
We remark that conclusions about the connectedness based on state evolution are not possible, since the adiabatic theorem does not hold for the non-Hermitian models discussed here~\cite{Nenciu1992,Uzdin2011,Milburn2015,Zhang2025}.
Prominent platforms providing the necessary tunability are optical or plasmonic metasurfaces, which rely on artificial units cells defined in periodic arrays of nanoantennas~\cite{Ozdemir2019,Opala2023}. 
Loss generates non-Hermitian contributions to the dynamics of these systems.
Such setups are capable of realizing the two-band model $H_\text{FC}(\bm{k})$ in a two-orbital square lattice given full control over the onsite terms and hoppings up to next-nearest-neighbor distance.
Mechanical metamaterials, in which individual oscillators are driven depending on the state of the system, or single-photon interferometry, recently established to study non-Hermitian topological features~\cite{Wang2024}, provide other feasible platforms for the observation of EP pair creation and their threading through the Brillouin zone. They are thus capable of resolving the transition between topologically distinct states.

We briefly outline how to implement the model defined in Eq.~(\ref{eq:nH_toric}) on a non-Hermitian acoustic metasurface (phononic crystal)~\cite{Liu2022} and how to detect the Fermi cuts.
Consider an acoustic metasurface given by a two-dimensional square lattice with two separate metal cavities per unit cell.
The height of the cavities is adjustable, realizing on-site potentials; nearest and next-nearest neighbor cavities are connectable by tunable channels, resulting in tight-binding hopping terms; and both cavities and channels are outfitted with speakers allowing for adjustable non-Hermitian contributions.
The setup is controlled by measuring amplitude and phase of the standing waves within the cavities and precisely controlling the feedback via the speakers.
Such an implementation of non-Hermitian lattices was recently realized, demonstrating that access to their spectral response is experimentally feasible~\cite{Liu2022}.
\begin{figure}
    \centering
    \includegraphics[width=\linewidth]{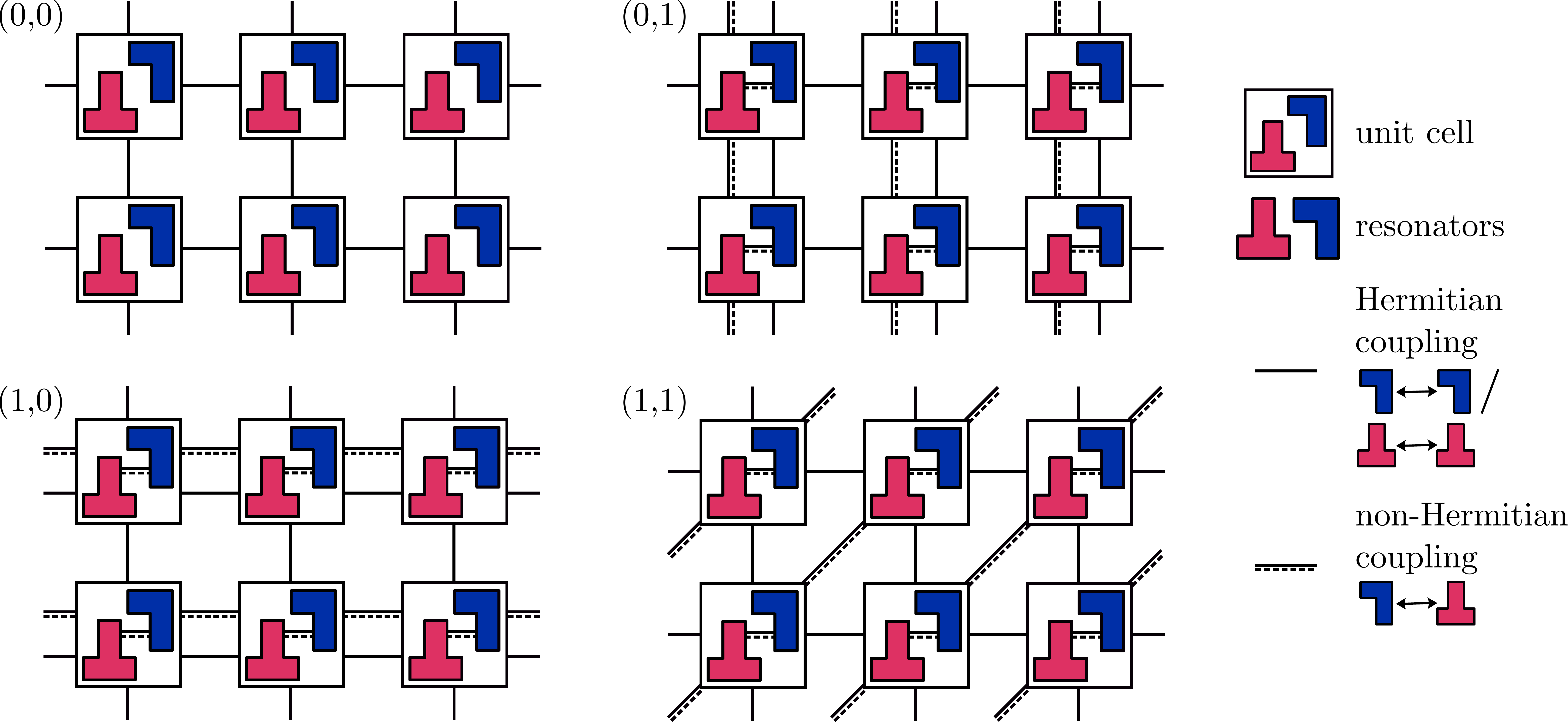}
    \caption{Schematic configuration of the coupling structure in an non-Hermitian acoustic metasurface to realize topologically distinct Riemann energy sheets. Tuning the values of the non-Hermitian couplings by adjusting the feedback of the speakers implements the Bloch Hamiltonians $H_\text{FC}^{(m_x,m_y)}(\bm{k})$ in Eq.~(\ref{eq:nH_toric}). Transitions between topologically distinct Riemann sheet configurations can be measured by adjusting multiple non-Hermitian couplings simultaneously.}
    \label{fig:acoustic_metasurface}
\end{figure}
In Figure~\ref{fig:acoustic_metasurface} different acoustic metasurface are shown, which realize the topologically distinct Riemann sheet configurations. 
By detecting the pressure response of an acoustic source placed in the lattice and Fourier transforming the spatial data into momentum space, the dispersion relation can be measured.
This dispersion corresponds to the real part of the complex-valued spectrum and thus the presence or absence of non-contractible Fermi cuts in the Riemann energy sheets is accessed.

\section{Conclusion}
We have demonstrated topological structures tied to the presence of an energy gap in the spectral Riemann sheets of non-Hermitian systems.
Pairs of EPs emerge as uniquely non-Hermitian features in these complex-valued energy structures. 
We introduce Fermi cuts as the branch cuts connecting EPs along lines with degenerate real energy parts.
These cuts can be closed non-trivially by separating the EP pair, and recombining it across the Brillouin-zone boundary.
The EPs then annihilate, leaving a non-contractible closed Fermi cut.
Imposing time-reversal symmetry on the non-Hermitian system reduces the number of possible topologically distinct gapped Riemann energy sheet configurations, resulting in a characterization by a $\mathbb{Z}_2 \times \mathbb{Z}_2$ topological invariant.
Building on an analogy to Kitaev's toric code, these Fermi cuts resemble non-contractible closed defect lines on the toroidal Brillouin zone of a two-dimensional lattice.
Four topologically distinct configurations of the non-Hermitian system are characterized by a fully non-degenerate spectrum and EPs are interpreted as excitations.
Non-Hermitian multi-band systems facilitate multiple different excitations and the presence or absence of the closed Fermi cuts establishes topologically distinct configurations.
The energy spectrum behaves as a classical object, so that this analogy does not inherit the full quantum-mechanical properties of the toric code. 
It instead gives rise to Riemann sheet structures beyond the conventional toric-code dimension.
The resulting configurations are topologically protected and remain stable under perturbation, due to finite gaps in the complex-valued energy spectra.
This framework establishes a novel approach to utilize the topological features inherent in non-Hermitian systems.

\section*{Acknowledgements}
We are grateful to Joost Slingerland for his lectures during the Young Researchers School 2024 in Maynooth and to both J. Lukas K. K\"{o}nig and Lukas R{\o}dland for fruitful discussion.

\paragraph{Funding information}
A.M., A.F. and F.K.K acknowledge funding from the Max Planck Society's Lise Meitner Excellence Program 2.0.

\bibliography{references.bib}

\end{document}